\documentstyle[pra,aps,multicol,epsfig]{revtex}

\newcommand\be{\begin{eqnarray}}
\newcommand\ee{\end{eqnarray}}
\newcommand\ba{\begin{array}}
\newcommand\ea{\end{array}}
\def\r{\rangle}

\begin{document}
\title{Instability and entanglement of the ground state of the
Dicke model}
\author{Vladim\'{\i}r Bu\v zek$^{1,2}$, Miguel Orszag$^{3}$, and Mari\'{a}n Ro\v{s}ko$^{1,2}$}
\address{$^{1}$
Research Center for Quantum Information, Slovak Academy
of Sciences,
D\'ubravsk\'a cesta 9, 845 11 Bratislava, Slovakia \\
${}^{2}$ {\em Quniverse}, L{\'\i}\v{s}\v{c}ie \'{u}dolie 116, 841 04 Bratislava, Slovakia
\\
${}^{3}$ Facultad de F\'{\i}sica, Pontificia Universidad
Cat\'{o}lica de Chile, Casilla 306, Santiago, Chile
}

\date{1 February 2005}
\maketitle

\begin{abstract}
Using tools of quantum information theory we show that the ground
state of the Dicke model exhibits an infinite sequence of
instabilities (quantum-phase-like transitions). These transitions
are characterized by abrupt changes of the bi-partite entanglement between atoms
at critical values $\kappa_j$ of the atom-field coupling parameter $\kappa$
and are accompanied by discontinuities of the first derivative of the energy
of the ground state. We show that in a weak-coupling limit
($\kappa_1\leq \kappa \leq \kappa_2$) the Coffman-Kundu-Wootters (CKW) inequalities are
saturated which proves that for these values of the coupling no intrinsic
multipartite entanglement (neither among the atoms nor between the atoms and the field)
is generated by the atom-field interaction. We analyze also the atom-field entanglement and
we show that in the strong-coupling limit the field is entangled with the atoms
so that the von Neumann entropy of the atomic sample (that serves as a measure of the atom-field entanglement) takes the
value $S_A=\frac{1}{2}\ln (N+1)$. The entangling interaction with atoms leads to a highly
sub-Poissonian photon statistics of the field mode.
\end{abstract}


\begin{multicols}{2}

Interactions between quantum objects lead to correlations
that may have no classical analogue. These purely quantum
correlations, known as entanglement, play fundamental role in
modern physics and have already found their applications in
quantum information processing and communication
\cite{Nielsen2000}. A degree of quantum entanglement depends on the
physical nature of interacting objects and on the character of
their mutual coupling.  There is another physical phenomenon that
shares many of the features with the entanglement - quantum phase
transitions (QPT) \cite{Sachdev2000}. These transitions occur at
zero temperature and they are induced by the change of a
coupling parameter. Recently several authors have reported on a
close relation between quantum critical phenomena and appearance
of entanglement. In particular, scaling of entanglement close to a
quantum phase transition in an Ising model has been reported
\cite{Osterloh2002}.
In this Letter we report on an infinite sequence of instabilities
(i.e., a sequence of quantum-phase-like transitions) and its connection
to {\em bi-partite} atomic entanglement in
the so-called Dicke model (DM) \cite{Dicke1954,Tavis1967,Narducci1973,Andreev1993,Benedict1996}.
The  QPT in the DM was first rigorously
described by Hepp and Lieb \cite{Hepp1973} (see also Ref.~\cite{Wang1973})
while the instability of
the ground state has been reported by Narducci {\it et al.} \cite{Narducci1973}.
In the paper by Hepp and Lieb \cite{Hepp1973} only a thermodynamics limit of  large-$N$ and weak-coupling
$\kappa$ has been studied. It has been shown that in this limit at $\kappa=\omega/\sqrt{N}$ the ground state of the model
exhibits the second-order QPT.
QPT in the DM have been studied by Hillery and Mlodinow \cite{Hillery1984}
and Rzazewski {\it et al.} \cite{Rzazewski1975} in the limit of
large $N$ and weak coupling.
QPT
and its relation to a chaos have been studied  by Emary and Brandes \cite{Emary2003}.
Schneider and Milburn \cite{Schneider2002} (see also
Refs.~\cite{Messikh2003,Basharov2003}) studied appearance of entanglement and its connection to quantum phase transitions in the DM
with just two atoms. The connection between the atom-field entanglement and the QPT in the thermodynamical limit of the DM
has recently been studied by Lambert {\it et al.} \cite{Lambert2004}.
However, in all these studies the {\it sequence} of ground-state instabilities (quantum-phase-like transitions)
and the behavior of bi-partite atomic concurrence have been completely overlooked.
In this Letter we report on this remarkable feature of quantum entanglement in the Dicke model
both in the weak-coupling as well as strong coupling regimes.

Let us assume a model of $N$ identical two-level atoms (spin-1/2 particles), at positions
$\vec{r}_1,\dots,\vec{r}_N$ coupled to a single mode electromagnetic field via the electric-dipole
interaction originally proposed by Dicke \cite{Dicke1954}
(see also Refs.~\cite{Tavis1967,Narducci1973,Andreev1993,Benedict1996}).
In the rotating-wave approximation the model Hamiltonian reads
\be
\label{2.1}
H&=&\frac{\hbar \omega_A}{2}\sum_{j=1}^N \sigma_j^z + \hbar \omega_F a^\dagger a
\nonumber
\\
&+& \hbar \kappa
\sum_{j=1}^N
\left({\rm e}^{i \vec{r}_j\cdot \vec{k}_j} a^\dagger \sigma_j^-
+ {\rm e}^{-i \vec{r}_j\cdot \vec{k}_j} a \sigma_j^+\right)\, ,
\ee
where $a$ and $a^\dagger$ are the field annihilation and creation operators, respectively.
In what follows we will
assume that the atomic frequency $\omega_A$ is resonant with the
field frequency $\omega_F$ and $\hbar=1$. All the atoms are assumed
to be coupled to the field with the same coupling  $\kappa$ and the position dependent phase factors
can be included into definition of the Pauli matrices $\sigma$ via local unitary transformations (see e.g.
Ref.~\cite{Hillery1984}).
As shown by Tavis and Cummings \cite{Tavis1967} one of the integrals of motion of the DM
corresponding to the total excitation number
$P=a^\dagger a + \sum_{j=1}^N \sigma^+_j \sigma^-_j$
partitions the total Hilbert space of the DM into a direct sum of subspaces labelled by different
excitations numbers.

When the number of excitations $p$ is smaller than the number
of atoms $N$, then the corresponding subspace of the Hilbert space is spanned by $p+1$ vectors $|\{e^s,g^{\otimes (N-s)}\}\r_A|p-s\r_F$
with $s=0,\dots,p$ where $|\{e^s,g^{\otimes (N-s)}\}\r_A$ is a completely symmetric state of $s$ atoms in the upper level $|e\r$ and
$(N-s)$ atoms in the lower level $|g\r$, while $|p-s\r_F$ describes a Fock state with $(p-s)$ photons.
 The DM Hamiltonian (\ref{2.1}) in this subspace is represented by a 3-diagonal $(p+1)\times(p+1)$ matrix the eigenvalues
$E_j^{(p)}(\kappa)$ of which are linear functions of the coupling $\kappa$, i.e. $E_s^{(p)}(\kappa)=K_s^{(p)}\kappa + E_0^{(p)}$, where
$E_0^{(p)}=(p-N/2)\omega$ and $K_s^{(p)}$ are parameters to be determined from the eigenvalue problem. The lowest energy eigenstate in the
given subspace of the Hilbert space is determined by the smallest (negative) $K_{min}^{(p)}$. In what follows we will omit the subscript {\it min}
and when we use $E^{(p)}(\kappa)$ and $K^{(p)}$ we refer to the energy of the ground state for a given value of $\kappa$.


The character of the ground state of the DM (\ref{2.1}) with the smallest eigenenergy,
depends on the coupling  $\kappa$. At the critical values of the coupling  $\kappa_j$
the number of excitations $p$ in the ground state of the system, the bi-partite concurrence $C$ between atoms as well as
the von Neumann entropy of the field mode do exhibit discontinuities. These effects can be clearly seen from Table~1, and
Figs.~1 and 2. Specifically, let us consider eigenenergies of the ground state with 0 and 1 excitations. Their explicit expressions
are $E^{(0)}=-\frac{N}{2}\hbar\omega$ and $E^{(1)}=\frac{2-N}{2}\omega -\kappa \sqrt{N}$. It is clear that for $\kappa/\omega<1/\sqrt{N}$ the
energy $E^{(0)}$ is smaller than $E^{(1)}$, so the state $|g^{\otimes N}\r$ with zero excitations is the ground state of the DM. But for
$\kappa/\omega> 1/\sqrt{N}$ this state is not a ground state of the DM anymore since $E^{(1)}<E^{(0)}$. That is $\kappa_1=\omega /\sqrt{N}$ is the
critical point at which number of excitations in the ground state of the DM is abruptly changed. Analyzing the inequalities between
eigenenergies of lowest energy eigenstates in subspaces of the Hilbert space with different excitations numbers we can determine
critical values of the coupling parameter $\kappa$.
The first few critical values can be expressed in a simple analytical form
\be
\kappa_1/\omega & = & N^{-1/2}\, ;
\\
\kappa_2/\omega &=&[\sqrt{4N-2} - \sqrt{N}]^{-1}\, ;
\nonumber
\\
\kappa_3/\omega&=&[\sqrt{5(N-1) + \sqrt{(4N-5)^2 + 8N}} - \sqrt{4N-2}]^{-1}\, .
\nonumber
\ee
In a thermodynamical limit of large $N$
(while $p\ll N$) all critical points ``merge'' together, i.e. $\kappa_j\simeq \omega/\sqrt{N}$ (where $j=1,\dots,p$) and we recover the
result of Hepp and Lieb \cite{Hepp1973}.

On the other hand in the strong-coupling regime ($p\geq N$)
when the Hamiltonian is represented by the $(N+1)\times (N+1)$ matrix (the size of which does not depend on the particular value of $p$)
the energy of the ground state of the DM again linearly depends on $\kappa$ and for a given $p\gg N$ it can be approximated as
$E^{(p)}=-N \sqrt{p} \kappa + p\omega$. The critical values of $\kappa$
in the limit $p\gg N$ can be approximated as $\kappa_p\simeq \omega [N(\sqrt{p+1}-\sqrt{p})]^{-1}$.

In our approach we consider distinguishable atoms (located at positions $\vec{r}_1,\dots,\vec{r}_N$)
that interact with a single mode field. This field mediates quantum correlations between the atoms.
In what follows we will study bi-partite atomic entanglement
in the ground state of the Dicke system
that for a given $\kappa$ is represented by a superposition
$
|\Xi^{(p)}\r=\sum_{s=0}^p A_s |\{e^{\otimes s};g^{\otimes(N-s)}\}\r_A |p-s\r_F\, .
$
The composite states $|\{e^{\otimes s};g^{\otimes(N-s)}\}\r$ of this superposition exhibit entanglement.
For instance the state $|\{e;g^{\otimes(N-1)}\}\r$ is the state (the so-called entangled web)
with maximal bi-partite entanglement between an
arbitrary pair of atoms (see Koashi {\it et al.} \cite{Koashi2000}).

\begin{figure}
\centerline{
\includegraphics[width=8cm]{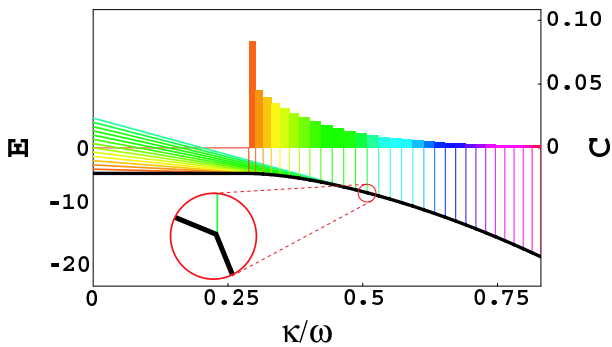}}
\bigskip
\caption{ The energy $E$ (measured in units of $\hbar\omega$) of the ground state of the DM with $N=12$
atoms as a function of the scaled coupling parameter $\kappa/\omega$. We see
a sequence of instabilities (quantum-phase-like transitions), that are marked
 by vertical lines corresponding to different values of the coupling
$\kappa_j$. The first interval of values of the coupling parameter $0\leq \kappa \leq \kappa_1$
is characterized by the excitation number equal to zero. The second interval
$\kappa_1\leq \kappa \leq \kappa_2$ is characterized by one excitation, etc.
The first derivative of the ground-state energy (black line) is not a
continuous function of $\kappa$ (see the inset that illustrates the behavior of the energy of the ground state around the critical point $\kappa_j$).
 In the figure we plot also a bi-partite concurrence as a function of $\kappa$.
The concurrence exhibits abrupt changes at ``critical'' values of $\kappa_j$.
The maximal value of bi-partite concurrence in the Dicke system appears for
$\kappa_1\leq \kappa \leq \kappa_2$ when the ground state of the system contains just one excitation ($p=1$).
Then, the value of the concurrence decrease with the increase of $p$. Nevertheless, even for $p\gg 12$ the concurrence is non-zero.
} \label{fig1}
\end{figure}

\begin{figure}
\centerline{
\includegraphics[width=6cm]{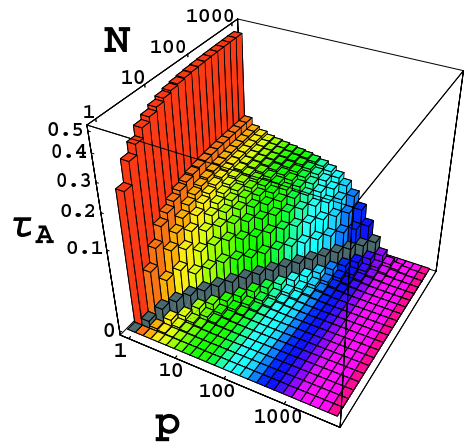}}
\bigskip
\caption{
The bi-partite concurrence of the ground state of the DM as a function of number
of atoms $N$ and the number of excitations $p$.
Since the bi-partite concurrence decreases as $1/N$
we plot the ``total atomic bi-partite entanglement'' $\tau_A=C^2 N(N-1)/2$,
where the factor $N(N-1)/2$ corresponds to the number of possible pairs in the atomic
sample. We see that $\tau_A$ as a function of $N$ takes the largest value
for $p=1$ and in the large $N$ limit tends to the value 1/2.
We see that even in the strong coupling case when $p=N$  arbitrary pair of atoms in the sample is entangled. This is reflected
by non-zero value of $\tau_A$ (see the diagonal line in the $(p,N)$ plane of the figure). In this case, for $N$ large enough the field
exhibits sub-Poissonian photon statistics with the mean-photon number $\bar{n}=2N/3$ and the dispersion $\Delta=\sqrt{5N/26}$.
 We conclude that larger the number of excitations $p$ smaller is the amount of bi-partite entanglement
between atoms. In the strong-coupling limit a strong atom-field entanglement is established that is quantified by the von-Neumann entropy
$S_A=S_F=\frac{1}{2}\ln (N+1)$.
} \label{fig2}
\end{figure}

In order to make our discussion  quantitative we recall the definition of the concurrence \cite{Hill1997,Wootters1998}
between a pair of qubits (two-level atoms). Let $\varrho$ is a density matrix of the pair of two-level atoms
expressed in the basis $\{|gg\r, |ge\r, |eg\r, |ee\r\}$. Let $\tilde{\varrho}$ is a matrix defined as
$\tilde{\varrho}=(\sigma^y\otimes\sigma^y)\varrho^T (\sigma^y\otimes\sigma^y)$, where $\sigma^y$ is a Pauli matrix
$\left(\begin{array}{cc} 0 & -i \\ i & 0\end{array}\right)$, while $T$ indicates a transposition. Then the concurrence
of a bi-partite system  reads $C= {\rm max}\{\lambda_1 - \lambda_2
-\lambda_3-\lambda_4, 0\}$, where $\lambda_i$ are the square roots of the eigenvalues of the matrix
$\tilde{\varrho}\varrho$
in descending order.  The values of concurrence range from zero (for
separable states), to one (for maximally entangled states of two qubits).
The bipartite entanglement of the entangled web state $|\{e;g^{\otimes(N-1)}\}\r$
measured in concurrence takes the value $C=2/N$.
As follows from our investigations (see Table I and Fig.~1) for $\kappa_1\leq \kappa\leq\kappa_2$ the entanglement
between an arbitrary pair of atoms corresponds to the concurrence $C^{(1)}=1/N$. Increasing the coupling
the value of the concurrence is decreasing but it changes {\em discontinuously} at critical values  $\kappa_j$ (see Fig.~\ref{fig1}).
Nevertheless,
the total amount of the bipartite entanglement in the atomic sample measured in terms of the function $\tau_A$ is non-zero
for arbitrarily large number $N$ of atoms and arbitrarily large coupling  (see Fig.~\ref{fig2}).

The bi-partite entanglement between atoms is mediated via a single-mode electromagnetic field that interacts with the atoms
in the dipole and the rotating-wave approximations. In fact, this interaction leads also to an entanglement of the field mode
to the atomic system. In order to measure the degree of entanglement between the field and the atoms
 we  utilize the von Neumann entropy of the atomic sample $S_A=-{\rm Tr} [\rho_A \ln \rho_A]$,
where $\rho_A$ is the density operator of the atoms in the ground state $|\Xi^{(p)}\r$. It follows from the Araki-Lieb theorem
that for a pure atom-field state $|\Xi^{(p)}\r_{AF}$ the field entropy $S_F$ is equal to $S_A$.
Larger the entropy is the larger the degree of entanglement.
For $p\leq N$ the upper bound on the entropy $S_A$ is given by the number of excitations $p$, i.e. $S_A^{(max)}=\ln (p+1)$, while for
$p\geq N$ the uppper bound is given by the total number of atoms in the system, i.e. $S_A^{(max)}=\ln (N+1)$.
For a given number of atoms the atomic (field) entropy depends on the number of excitations in ground state of the DM.
 For very weak coupling, i.e.
when $0\geq \kappa \geq \kappa_1$ the atoms are totally disentangled from the field.
In the region of values
$\kappa_1\leq\kappa\leq\kappa_2$ when the ground state is characterized by one excitation, the field mode occupies a Hilbert space
spanned by two Fock states $|0\rangle_F$ and $|1\rangle_F$. In this case, the entropy is maximal and equals to
$\ln 2$. With the increase of the coupling  $\kappa$ the entropy of the field is always smaller than $\ln (p+1)$. In fact,
for large $N$ and for $p$ large ( $p\geq N$)
we find that $S_A=\frac{1}{2} \ln(N+1)$, i.e. the entropy is equal to  {\it one half} of its maximally possible value.
That is the field is not maximally entangled
with the atomic sample. The reason for this behavior is that the entanglement cannot be shared freely. To illuminate this
issue let us consider the range of $\kappa$ such that $p=1$. In this case the field mode can be effectively  represented
as a qubit. Consequently, the system of $N$ two-level atoms and the field mode can be represented as a set of
$(N+1)$ qubits. Recently Coffman {\em et al.} \cite{Coffman2000} have conjectured that for pure states the sum of square of
concurrencies between a qubit $j$ and any other qubit $k$ is smaller than or equal to the tangle between the given qubit $j$
and the rest of the system, i.e. the CKW inequalities read
\be
\sum_{k=1;k\neq j}^N C^2_{j,k}\leq \tau_{j,\overline{j}}=4 {\rm det}\rho_j\, ,
\ee
where the sum in the left-hand-side is taken over all qubits
except the qubit $j$, while $\tau_{j,\overline{j}}$ denotes the
tangle between the qubit $j$ and the rest of the system
(denoted as $\overline{j}$).
If we assume that the qubit $j$ represents the field mode, then we can find that the tangle between the field
and the system of atoms reads $\tau_{FA}=4 {\rm det} \rho_F=1$, while the concurrencies between the field and each of the atoms is
$C_{FA_j}=1/\sqrt{N}$.
From here it directly follows that
the ground state of the DM with $p=1$ saturates the CKW inequalities, which proves that the
atom-field interaction as described by the Hamiltonian (\ref{2.1})
with small coupling ($\kappa_1\leq \kappa\leq \kappa_2$)
does induce only {\em bi-partite} entanglement and does not result in intrinsic multipartite quantum correlations.
 For the moment it is essentially impossible to generalize
this result for other values of $p$ since no measures of entanglement between a qudit (field mode with $p>1$ excitations)
and a set of qubits (atoms) are available and no corresponding generalization of the CKW inequalities is known.
Nevertheless, it is interesting to study the strong-coupling limit when $p\gg N$ from a different perspective.
In this case the ground state of DM
reads
$
|\Xi^{(p)}\r_{AF}=\sum_{s=0}^N A_s |\{e^{\otimes s};g^{\otimes(N-s)}\}\r_A |p-s\r_F \, .
$
For $N\ll p$ the amplitudes $A_s$ can be approximated as
$A_s\simeq (-1)^{N+s}\sqrt{\left(\begin{array}{c} N\\s \end{array}\right) 2^{-N}}$.
We already know that in this case
the field mode is highly entangled with the atoms that is reflected by the field entropy
$S_F\simeq \frac{1}{2}\ln (N+1)$.
Quantum correlations that are established in the atom-field system  lead to highly non-trivial photon statistics of the field mode.
Specifically, from above it follows that for $p\gg N$ the mean excitation of the atomic sample is equal to $N/2$. Correspondingly,
the mean photon number of the field mode is equal to $\bar{n}=p-N/2$ with the dispersion of the photon number distribution equal to
$\Delta = \sqrt{N}/2$. This means that in the strong-coupling limit
entanglement between atoms and the field leads to highly sub-Poissonian photon statistics of the field mode in the ground state of the DM.
This is reflected by the Mandel parameter of the field mode, defined as $Q=(\Delta -\bar{n})/\bar{n}$, that is close to -1 for $p\gg N$.

In summary, using tools of quantum information theory we have shown that the ground
state of the Dicke model exhibits an infinite sequence of
instabilities (quantum-phase-like transitions). These transitions
are characterized by abrupt changes of the bi-partite entanglement between atoms
at critical values $\kappa_j$ of the atom-field coupling parameter $\kappa$.

 This was work supported in part by  the European Union project
CONQUEST and by the Slovak Academy of Sciences via the project
CE-PI and the project APVT-99-012304. One of us (M.O.) would like
to thank Fondecyt (projects 1010777 and 7010777) and Milenio
  ICM (Po2-049) for partial support.

\end{multicols}
\begin{table}
\begin{tabular}{||l|l|l|l||}
$p$ & $\kappa/\omega$ & $E^{(p)}$ & $C^{(p)}$\\
\hline
0 & $0 \leq \kappa \leq \kappa _{1}$ & $E^{(0)}=-\frac{N}{2}\omega $ & $C^{(0)}=0$
\\
\hline\hline
1 & $\kappa _{1}\leq \kappa \leq \kappa _{2}$ &  $E^{(1)}=\frac{2-N}{2}\omega -\kappa
\sqrt{N}$ & $C^{(1)}=\frac{1}{N}$ \\
\hline
2 & $\kappa _{2}\leq \kappa \leq \kappa _{3}$ & $E^{(2)}=
\frac{4-N}{2}\omega -\kappa \sqrt{2(2N-1)}$ & $C^{(2)}=\frac{4N-5-2\sqrt{2N^{2}-5N+4}}{N(2N-1)}$ \\
\hline
3& $\kappa _{3}\leq \kappa \leq \kappa _{4}$ &
$E^{(3)}=\frac{6-N}{2}\omega -\kappa \sqrt{5(N-1)+\sqrt{(4N-5)^{2}+8N}}$ &
$\cdots$ \\\hline
4 & $\kappa _{4}\leq \kappa \leq \kappa _{5}$ &  $E^{(4)}=
\frac{8-N}{2}\omega -\kappa \sqrt{10N-15+3\sqrt{17-12N+4N^{2}}}$ & $\cdots$
\\
\end{tabular}
\smallskip
\caption{The number of excitations $p$ in the ground state of the DM with $N$ atoms changes abruptly at critical values
$\kappa_j$. The larger the coupling is the larger the number of excitations in the ground state.
The energy of the ground state $E$ is continuously decreasing with the increase of the coupling $\kappa$. For different regions of
$\kappa$ characterized by different excitation numbers the ground-state energy is presented
in Table~1 and in Fig.~1. At critical values $\kappa_j$ the first derivatives of the energy $\partial E/\partial \kappa$ exhibit
discontinuities while the bi-partite concurrence reflecting entanglement between pairs of atoms is changed abruptly (discontinuously).
}
\end{table}
\begin{multicols}{2}

\end{multicols}
\end{document}